\documentclass[10pt,prl,twocolumn,showpacs]{revtex4}
\usepackage{graphicx,amsmath,amssymb}
\begin{document}

\def\nn{\nonumber}
\def\kc#1{\left(#1\right)}
\def\kd#1{\left[#1\right]}
\def\ke#1{\left\{#1\right\}}
\renewcommand{\Re}{\mathop{\mathrm{Re}}}
\renewcommand{\Im}{\mathop{\mathrm{Im}}}
\renewcommand{\b}[1]{\mathbf{#1}}
\renewcommand{\c}[1]{\mathcal{#1}}
\renewcommand{\u}{\uparrow}
\renewcommand{\d}{\downarrow}
\newcommand{\bsigma}{\boldsymbol{\sigma}}
\newcommand{\blambda}{\boldsymbol{\lambda}}
\newcommand{\Tr}{\mathop{\mathrm{Tr}}}
\newcommand{\sgn}{\mathop{\mathrm{sgn}}}
\newcommand{\sech}{\mathop{\mathrm{sech}}}
\newcommand{\diag}{\mathop{\mathrm{diag}}}
\newcommand{\Pf}{\mathop{\mathrm{Pf}}}
\newcommand{\half}{{\textstyle\frac{1}{2}}}
\newcommand{\sh}{{\textstyle{\frac{1}{2}}}}
\newcommand{\ish}{{\textstyle{\frac{i}{2}}}}
\newcommand{\thf}{{\textstyle{\frac{3}{2}}}}
\newcommand{\SUN}{SU(\mathcal{N})}
\newcommand{\N}{\mathcal{N}}

\title{Fractional topological insulators in three dimensions}

\author{Joseph Maciejko$^1$, Xiao-Liang Qi$^{2,1}$, Andreas Karch$^3$ and Shou-Cheng Zhang$^1$}

\affiliation{$^1$Department of Physics, Stanford
University, Stanford, CA 94305, USA\\
$^2$Microsoft Research, Station Q, Elings Hall, University of
California, Santa Barbara, CA 93106, USA\\
$^3$Department of Physics, University of Washington, Seattle, WA
98195-1560, USA}

\date\today

\begin{abstract}
Topological insulators can be generally defined by a topological
field theory with an axion angle $\theta$ of $0$ or $\pi$. In this
work, we introduce the concept of fractional topological insulator
defined by a fractional axion angle and show that it can be
consistent with time reversal ($T$) invariance if ground state
degeneracies are present. The fractional axion angle can be
measured experimentally by the quantized fractional bulk
magnetoelectric polarization $P_3$, and a `halved' fractional
quantum Hall effect on the surface with Hall conductance of the
form $\sigma_H=\frac{p}{q}\frac{e^2}{2h}$ with $p,q$ odd. In the
simplest of these states the electron behaves as a bound state of
three fractionally charged `quarks' coupled to a deconfined non-Abelian
$SU(3)$ `color' gauge field, where the fractional charge of the
quarks changes the quantization condition of $P_3$ and allows
fractional values consistent with $T$-invariance.
\end{abstract}


\maketitle

Most states of quantum matter are classified by the symmetries
they break. However, topological states of quantum
matter~\cite{Qi2010} evade traditional symmetry-breaking
classification schemes, and are rather described by topological
field theories (TFT) in the low-energy limit. For the quantum Hall
effect, the TFT is the $2+1$ dimensional Chern-Simons (CS)
theory~\cite{Zhang1992} with coefficient given by the quantized
Hall conductance. In the noninteracting limit, the integer
quantized Hall (IQH) conductance in units of $\frac{e^2}{h}$ is
given by the TKNN invariant~\cite{TKNN1982} or first Chern number.
In the presence of strong correlations, one can also observe the
fractional quantum Hall effect (FQHE), where the Hall conductance
is quantized in rational multiples of $\frac{e^2}{h}$. In both
cases however, these topological states can exist only in a strong
magnetic field which breaks time reversal ($T$) symmetry.

More recently, $T$-invariant topological insulators (TI) have been
studied extensively~\cite{Qi2010,Moore2009,Hasan2010}. The TI
state was first predicted theoretically in HgTe quantum wells, and
observed experimentally~\cite{bernevig2006d,koenig2007b,Kane2005,bernevig2006a} soon
after. The theory of TI has been developed along two independent
routes. Topological band theory identified $\mathbb{Z}_2$
topological invariants for noninteracting band
insulators~\cite{Kane2005,Fu2007,Moore2007}. The TFT of
$T$-invariant insulators was first developed in $4+1$ dimensions,
where the CS term is naturally
$T$-invariant~\cite{Zhang2001,Bernevig2002}. Dimensional reduction
then gives the TFT for TI in $3+1$ and $2+1$
dimensions~\cite{Qi2008}. The TFT is generally valid for
interacting systems, and describes the experimentally measurable
quantized magnetoelectric response. The coefficient of the
topological term, the axion angle $\theta$, is constrained to be
either $0$ or $\pi$ by $T$-invariance. The TFT has been further
developed in Ref.~\cite{Essin2009,Karch2009}. More recently, it
has been shown that it reduces to the topological band theory in
the noninteracting limit~\cite{Wang2009}.

By analogy with the relation between the IQHE and FQHE, one is
naturally led to the question whether there can exist a
`fractional TI'. In $2+1$ dimensions, an explicit wavefunction for
the fractional quantum spin Hall state was first proposed in
Ref.~\cite{bernevig2006a}, and the edge theory was investigated in
Ref.~\cite{Levin2009b}. The $T$-invariant fractional topological
state has also been constructed explicitly in $4+1$
dimensions~\cite{Zhang2001}. Since $T$-invariant TI form a
dimensional ladder in $4$, $3$ and $2$
dimensions~\cite{Qi2008,Schnyder2008,Kitaev2009}, it is natural to
investigate the $T$-invariant TI in $3+1$ dimensions. Fractional
states generally arise from strong interactions. Since topological
band theory cannot describe such interactions, we formulate the
general theory in terms of the TFT. The TI is generally described
by the effective action
$S_\theta=\frac{\theta}{2\pi}\frac{e^2}{2\pi}\int
d^3x\,dt\,\b{E}\cdot\b{B}$ where
$\b{E}$ and $\b{B}$ are the electromagnetic fields~\cite{Qi2008}.
Under periodic boundary conditions, the partition function and all
physical quantities are invariant under shifts of $\theta$ by
multiples of $2\pi$. Since $\b{E}\cdot\b{B}$ is odd under $T$, it
appears that the only values of $\theta$ allowed by $T$ are $0$ or
$\pi$ mod $2\pi$.

In this paper, we show that there exist $T$-invariant insulating
states in $3+1$ dimensions with $P_3\equiv \frac{\theta}{2\pi}$ quantized in non-integer,
rational multiples of $\frac{1}{2}$ of the form
$P_3=\frac{1}{2}\frac{p}{q}$ with $p,q$ odd integers. The magnetoelectric polarization $P_3$ is defined by the response equation $\mathbf{P}=-\frac{\mathbf{B}}{2\pi}(P_3+\textrm{const.})$, where $\mathbf{B}$ is an applied magnetic field and $\mathbf{P}$ is the induced electric polarization. Such a
fractionalized bulk topological quantum number leads to a
fractional quantum Hall conductance of $\frac{p}{q}\frac{e^2}{2h}$
on the surface of the fractional TI. In contrast to the usual QHE
in $2+1$ dimensions, the surface QHE does not necessarily exhibit
edge states and thus cannot be directly probed by transport
measurements. Alternatively, it can only be experimentally
observed through probes which couple to each surface separately,
such as magneto-optical Kerr and Faraday
rotation~\cite{Qi2008,Karch2009}. Generically, a slab of
fractional TI can have different fractional Hall conductance on
the top and bottom surfaces, which can
be determined separately by combined Kerr and Faraday
measurements, independent of non-universal properties of the
material~\cite{maciejko2010}. Our approach is inspired by the
composite particle, or projective construction of FQH
states~\cite{Zhang1992,Jain1989,WenPartons,Barkeshli2009,Levin2009c}. The
idea is to decompose the electron with charge $e$ into $\N$
fractionally charged, fermionic `partons', which have a dynamics
of their own. One considers the case that the partons form a known
topological state, say a topological band insulator. When the
partons are recombined to form the physical electrons, a
new topological state of electrons emerges. In the FQH case for
example, the $\nu=\frac{1}{3}$ Laughlin state can be obtained by
splitting each electron into $\N=3$ partons of charge
$\frac{e}{3}$. Each parton fills the lowest Landau level and forms
a noninteracting $\nu=1$ IQH state. Ignoring the exponential factors, the parton wavefunction is the Slater determinant IQH
wavefunction
$\Psi(\{z_i\})\propto\prod_{i<j}\left(z_i-z_j\right)$, and the
electron wavefunction is obtained by gluing three partons
together, which leads to the Laughlin wavefunction
$\Psi_{1/3}(\{z_i\})\propto\prod_{i<j}\left(z_i-z_j\right)^3$.
Similarly, in 3+1 dimensions one can construct an interacting
many-body wavefunction by gluing partons which are in a
$\mathbb{Z}_2$ topological band insulator state. The parton ground
state wavefunction $\Psi_1(\{\b{r}_ns_n\})$ is a Slater
determinant describing the ground state of a noninteracting TI
Hamiltonian such as the lattice Dirac model~\cite{Qi2008}, with
$\{\b{r}_ns_n\}$, $n=1,\ldots,N$ the position and spin coordinates
of the partons. The electron wavefunction is obtained by requiring
the coordinates of all $\N_c$ partons
forming the same electron to be the same~\cite{Jain1989}, 
\begin{eqnarray}
\Psi_{\N_c}(\{\b{r}_ns_n\})
=\left[\Psi_1(\{\b{r}_ns_n\})\right]^{\N_c}.
\label{wavefunction3d}
\end{eqnarray}
Equation (\ref{wavefunction3d}) is the $(3+1)$-dimensional
generalization of the Laughlin wavefunction, and serves as a trial
wavefunction for the simplest fractional TI phases we propose.

\begin{figure}[t]
\begin{center}
\includegraphics[width=3.5in]{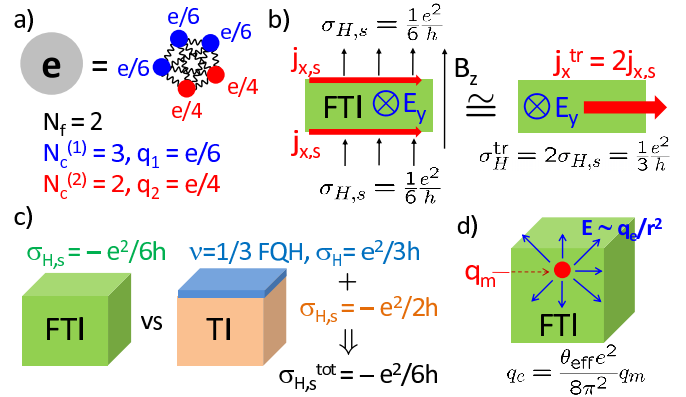}
\end{center}
\caption{a) Quark picture of fractional TI with flavor and color
degrees of freedom; b) surface FQHE vs transport measurements
[Eq.~(\ref{pq})]; c) nontrivial vs trivial fractional TI;
d) Witten effect as a probe of bulk topology.}
\label{fig:quarks}
\end{figure}

More generally, we can consider $\N_f$ different `flavors' of
partons, with $\N_c^{(f)}$ partons of each flavor
$f=1,\ldots,\N_f$. This decomposition has to satisfy two basic
rules. First, to preserve the fermionic nature of the electron,
the total number of partons per electron must be odd
(Fig.~\ref{fig:quarks}a),
\begin{align}\label{rule1}
\N_c^{(1)}+\N_c^{(2)}+\cdots+\N_c^{(\N_f)}=\text{odd}.
\end{align}
Second, if $q_f<e$ is the (fractional) charge of partons of flavor
$f$, the total charge of the partons must add up to the electron
charge $e$,
\begin{align}\label{rule2}
\N_c^{(1)}q_1+\N_c^{(2)}q_2+\cdots+\N_c^{(\N_f)}q_{\N_f}=e.
\end{align}
For instance, the $\nu=\frac{1}{3}$ Laughlin state described above
corresponds to $\N_f=1$, $\N_c^{(1)}=3$, and $q_1=\frac{e}{3}$,
which satisfies both conditions. Here we consider that partons of each flavor $f$ condense in a (generally
different) noninteracting $T$-invariant TI state with axion angle
$\theta_f=\pi\text{ mod }2\pi$. This is the analog of having
partons condense in various IQH states in the FQH construction.
Finally, the partons have to be bound together to yield physical
electrons. As we will see, this can be done by coupling
partons of flavor $f$ to a $SU(\N_c^{(f)})$ gauge field, which can be interpreted as a
`color field' where partons of flavor $f$ come in $\N_c^{(f)}$
colors. Since the TI analog of the $\nu=\frac{1}{3}$ Laughlin
state will involve three partons coupled to a $SU(3)$ gauge field
in $3+1$ dimensions, we dub our partons `quarks' by analogy with
quantum chromodynamics (QCD).

To obtain a more systematical understanding of the fractional TI,
we now deduce its effective gauge theory by way of a gedanken experiment. We consider subjecting a noninteracting TI to strong electron-electron interactions, and start with the simplest case of $\N_f=1$ with
$\N_c$ odd. The electron being split into $\N_c$ quarks of charge
$\frac{e}{\N_c}$, the electron operator will be written as a product of
$\N_c$ quark operators $\psi_{i\alpha}$, $i=1,\ldots,\N_c$.
However, the quark operators act in a Hilbert space which is
larger than the physical electron Hilbert space. We need to remove
those states of the quark Hilbert space which are not invariant
under unitary transformations which leave the electron
operator unchanged, i.e. $SU(\N_c)$ transformations with quarks in the $\N_c$ representation. The
projection onto the electron Hilbert space can therefore be
implemented by coupling the quarks to a $SU(\N_c)$ gauge
field $a_\mu$ with a coupling constant $g$. Outside the fractional TI, we expect the system to be in the confined phase, in analogy to quark confinement in QCD, which has only $SU(\N_c)$ singlet excitations in its low-energy spectrum, i.e. gauge-invariant `baryons'. Quarks of a given flavor within the baryon are antisymmetric in their ${\cal N}_c^{(f)}$ color indices; Fermi statistics then
implies that their spins are aligned. In a relativistic theory
this would imply that in the ${\cal N}_f=1$ theory the baryon has
spin $\frac{\N_c}{2}$. In nonrelativistic lattice models this is
not a concern, but even within the context of relativistic
continuum field theories one can obtain composite
spin-$\frac{1}{2}$ electrons for ${\cal N}_f>1$.

Inside the fractional TI, electron-electron interactions translate into complicated interactions among quarks. We consider the case that these interactions lead
the quarks to condense at low energies into a noninteracting
$T$-invariant TI state with axion angle $\theta$, and that the non-Abelian gauge field $a_\mu$ enters a deconfined phase~\cite{deconfined}. We now show that such a phase is a fractional TI. A low-energy effective
Lagrangian for $\N_f=1$ can be conjectured in the form
\begin{align}\label{Lquarks}
\mathcal{L}=\psi^\dag\left(iD_0-H_\theta[-i\b{D}]\right)\psi+
\mathcal{L}_\mathrm{int}(\psi^\dag,\psi),
\end{align}
where
$D_\mu=(D_0,-\b{D})=\partial_\mu+i\frac{e}{\N_c}A_\mu+iga_\mu$ is
the $U(1)_\mathrm{em}\times SU(\N_c)$ gauge covariant derivative,
and $H_\theta=H_\theta(\b{p})$ is the single-particle Hamiltonian
for a $T$-invariant TI with axion angle $\theta$.
$\mathcal{L}_\mathrm{int}$ represents weak $T$-invariant residual
interactions which do not destabilize the gapped TI phase, and can
thus be safely ignored. The kinetic Yang-Mills Lagrangian for
$a_\mu$ is generally present but not explicitly written.

Since the quarks are in a gapped TI phase, they can be integrated
out to yield an effective Lagrangian for the gauge fields~\cite{Qi2008},
\begin{align}\label{Leff}
\mathcal{L}_{\rm eff}&=\frac{\theta}{32\pi^2}\epsilon^{\mu\nu\lambda\rho}\Tr{\kc{
\frac{e}{\mathcal{N}_c}F_{\mu\nu}+gf_{\mu\nu}}\kc{\frac{e}{\mathcal{N}_c}F_{\lambda\rho}+gf_{\lambda\rho}}}\nn\\
&=\frac{\theta_\mathrm{eff}
e^2}{32\pi^2}\epsilon^{\mu\nu\lambda\rho}F_{\mu\nu}F_{\lambda\rho}+\frac{\theta
g^2}{32\pi^2}\epsilon^{\mu\nu\lambda\rho}\Tr
f_{\mu\nu}f_{\lambda\rho},
\end{align}
where $\Tr$ is the trace in the $\N_c$ representation of
$SU(\N_c)$, $F_{\mu\nu}=\partial_\mu A_\nu-\partial_\nu A_\mu$ and
$f_{\mu\nu}=\partial_\mu a_\nu-\partial_\nu a_\mu+ig[a_\mu,a_\nu]$
are the $U(1)_\mathrm{em}$ and $SU(\N_c)$ field strengths,
respectively, and the electromagnetic response is governed by an
effective axion angle
\begin{align}\label{fractheta}
\theta_\mathrm{eff}=\frac{\theta}{\N_c}=0,\pm\frac{\pi}{\N_c},
\pm\frac{3\pi}{\N_c},\pm\frac{5\pi}{\N_c},\ldots,
\hspace{5mm}\N_c\text{ odd}.
\end{align}
Equation~(\ref{Leff}) is obtained by replacing the
$U(1)_\mathrm{em}$ `electron' field strength $eF_{\mu\nu}$ in the
$U(1)_\mathrm{em}$ topological term
$\frac{\theta}{2\pi}\frac{e^2}{2\pi}\b{E}\cdot\b{B}= \frac{\theta
e^2}{32\pi^2}\epsilon^{\mu\nu\lambda\rho}
F_{\mu\nu}F_{\lambda\rho}$ for noninteracting TI~\cite{Qi2008} by
the total $U(1)_\mathrm{em}\times SU(\N_c)$ `quark' field strength
$\frac{e}{\N_c}F_{\mu\nu}+gf_{\mu\nu}$. Note that the crossed
terms of the
form $\Tr F_{\mu\nu}f_{\lambda\rho}$ vanish due to the tracelessness of the $SU(\N_c)$ gauge field. 
{More generally, $\theta_\mathrm{eff}$ can be obtained from the
Adler-Bell-Jackiw anomaly, since $\theta$ corresponds to the
phase of the quark mass~\cite{Qi2008}.} In principle, the
effective theory can also be obtained for quarks in a trivial
insulator state with $\theta=2n\pi,~n\in\mathbb{Z}$. However, such
a state is adiabatically connected to a trivial vacuum with
$\theta=0$, so that it is a trivial insulator in the bulk,
although a fractional $\theta_{\rm eff}$ can still be obtained due
to pure surface effects. Since the focus of the present work is a
fractional TI state with nontrivial bulk, we always consider
quarks with $\theta=\pi~{\rm mod}~2\pi$ in the following.

We are now faced with our initial question of whether the
effective theory~(\ref{Leff}),(\ref{fractheta}) breaks
$T$-invariance. According to the first term in Eq.~(\ref{Leff}),
$T$-invariance would require $\theta_\mathrm{eff}$ to be quantized
in integer multiples of $\pi$ if the minimal electric charge was
$e$~\cite{Witten1995}. However, the minimal charge in our theory
is $\frac{e}{\N_c}$, i.e. that of the quarks. Therefore,
$\theta_\mathrm{eff}$ has to be quantized in integer multiples of
$\frac{\pi}{\N_c^2}$. On the other hand, the
second term in Eq.~(\ref{Leff}) requires $\theta$ to be quantized
in integer multiples of $\pi$~\cite{langlands}, which means by
Eq.~(\ref{fractheta}) that $\theta_\mathrm{eff}$ has to be
quantized in units of $\frac{\pi}{\N_c}$. This latter constraint
is consistent with, but stronger than, the former~\cite{monopoles}, and the values
of $\theta_\mathrm{eff}$ allowed by $T$-invariance are thus
correctly given by Eq.~(\ref{fractheta}).

Equations~(\ref{Leff}) and~(\ref{fractheta}) constitute a TFT
which, precisely because it is topological, is insensitive to
small $T$-invariant perturbations and defines a new stable phase
of matter, the $T$-invariant fractional TI in $3+1$ dimensions.
The effective theory can also be derived in the multi-flavor case
$\N_f\geq1$, with $\N_c^{(f)}$ satisfying
rules~(\ref{rule1}) and (\ref{rule2}). Considering that quarks of
flavor $f$ form a noninteracting TI with axion angle
$\theta_f=\pi~{\rm mod}~2\pi$ and integrating them out yields an
effective Lagrangian in the form of~(\ref{Leff}), but with gauge
group $U(1)_\mathrm{em}\times
\prod_{f=1}^{N_f}U(\N_c^{(f)})/U(1)_\mathrm{diag}$. Here $U(1)_\mathrm{diag}$ is the overall $U(1)$ gauge transformation of the electron operator. 
The electromagnetic axion angle $\theta_\mathrm{eff}$ is given by
$\theta_\mathrm{eff}=\left(\sum_{f=1}^{\N_f}\frac{\N_c^{(f)}}{\theta_f}\right)^{-1}$.
When $\frac{\theta_f}{\pi}$ is odd for each flavor, one can show
that $\theta_{\rm eff}=\pi{p}/q$ with $p$, $q$ odd integers.

Important physical properties of the fractional TI can be read off from Eq.~(\ref{Leff}). The surface of the
fractional TI is an axion domain wall with the $U(1)_\mathrm{em}$
axion angle jumping from $\theta_{\rm eff}$ in the fractional TI
to $0$ in the vacuum. Such a domain wall has a surface QHE with
surface Hall conductance $\sigma_{H,s}=\frac{\theta_{\rm
eff}}{2\pi}\frac{e^2}{h}$~\cite{Qi2008}. Therefore, the surface Hall
conductance of the fractional TI has the general form
\begin{align}\label{pq}
\sigma_{H,s}=\frac{p}{q}\frac{e^2}{2h},
\hspace{5mm}p,q\text{ odd}.
\end{align}
For example, in the simplest single-flavor case with $\theta=\pi$
in Eq. (\ref{Leff}), we have
$\sigma_{H,s}=\frac{1}{\N_c}\frac{e^2}{2h}$ with $\N_c$ an odd
integer, corresponding to half of a $\nu=\frac{1}{\N_c}$ FQH
Laughlin state. The more general result (\ref{pq}) corresponds to
half of a generic Abelian
FQH state~\cite{FQHhierarchy,Jain1989}.

The fractional axion angle and the associated surface Hall
conductance~(\ref{pq}) are properties of the bulk topology. It is
important to distinguish them from a TI with
$\theta_\mathrm{eff}=\pm\pi$ and where the surface Dirac fermions
form a FQH state~\cite{ran2010}. In a noninteracting TI with
$\theta_\mathrm{eff}=-\pi$ for example, both the axion domain wall
and the surface FQH state contribute to $\sigma_{H,s}$,
\begin{align}
\sigma_{H,s}=\left(-\frac{1}{2}+\frac{n}{q}\right)\frac{e^2}{h}=\frac{2n-q}{q}\frac{e^2}{2h},
\end{align}
with $\frac{n}{q}$ an allowed filling fraction for a FQH state in
$2+1$ dimensions. For Abelian FQH states $q$ is odd, hence the
surface Hall conductance
$\sigma_{H,s}=\frac{2n-q}{q}\frac{e^2}{2h}$ has the same general
form as for the fractional TI [Eq.~(\ref{pq})]. As the simplest
example, the Laughlin state with $\frac{n}{q}=\frac{1}{3}$ leads
to $\sigma_{H,s}=-\frac16\frac{e^2}{h}$ (Fig.~\ref{fig:quarks}c,
right) which is the same as for a genuine fractional TI with bulk
$P_3=-\frac{1}{6}$ (Fig.~\ref{fig:quarks}c, left). However, the
bulk topology is very different in both cases. Therefore, surface
measurements are not sufficient to determine the bulk topology and
bulk measurements of $P_3$ are needed. One such measurement would
consist in embedding a monopole with magnetic charge $q_m$ inside
the fractional TI (Fig.~\ref{fig:quarks}d) and measuring its
electric charge $q_e$ induced by the Witten
effect~\cite{Witten1979,Qi2009,rosenberg2010}.

Another possible `experiment' is to measure the ground state
degeneracy (GSD) on topologically nontrivial spatial 3-manifolds. Consider for example a fractional TI on a manifold
$\Sigma_g\times\mathcal{I}$ with $\Sigma_g$ a Riemann surface of
genus $g$ and $\mathcal{I}=[0,L]$ a bounded interval, where $L$ is
the sample `thickness' and the two copies of $\Sigma_g$ (at each
end of ${\cal I}$) are the two bounding surfaces. We first discuss contributions to the GSD arising solely from the boundary, and comment on bulk contributions later on. A noninteracting
TI with a $\nu=\frac{1}{3}$ Laughlin state deposited on both
surfaces is described by two independent CS
theories~\cite{Zhang1992,topdegeneracy} and has a GSD of $3^g$
($m^g$ for $\nu=\frac{1}{m}$) on each surface for a total GSD of
$(3^g)^2=3^{2g}$. The situation is different for a genuine
fractional TI with $P_3=\pm\frac{1}{6}$ (Fig.~\ref{fig:quarks}c,
left). To study the GSD we set the external electromagnetic fields
to zero in Eq.~(\ref{Leff}) and consider the internal $SU(3)$
$\theta$-term. Assuming that the system stays gapped as we take
the limit of zero thickness $L\rightarrow 0$ where the gauge
fields $a_\mu$ on both surfaces become identified, the system is
described by a single $SU(3)_k$ CS theory on $\Sigma_g$ where the
level $k$ is the sum of the contributions from both surfaces. If
on both surfaces $\theta$ goes to the same value outside the TI,
then $k=0$ and there is no GSD. If $\theta=0$ on one side and
$\theta=2\pi$ on the other, we have a $SU(3)_1$ CS theory with GSD
$3^g\neq 3^{2g}$~\cite{topdegeneracy}. Another example is the
solid torus $S^1\times D^2$. The unique boundary is a 2-torus
$T^2$, hence a $\nu=\frac{1}{3}$ Laughlin state deposited on it
has GSD $3$ from the quantum mechanics of Wilson lines along the
two non-contractible loops on $T^2$~\cite{topdegeneracy}. However,
since one of these loops extends into the bulk of the solid torus
and is thus contractible, the $\theta$-term contributes no GSD to a genuine fractional TI. However, in addition to the boundary contributions of the $\theta$-term to the GSD, the gauge theory in the bulk can have a nontrivial GSD even in the absence of boundaries. For instance, the deconfined phase of $SU(\mathcal{N}_c)$ gauge theory has a GSD of $\mathcal{N}_c^3$ on $T^3$~\cite{sato2008}. The total GSD has in general both bulk and boundary contributions, and depends on the details of the gauge group.

It should be noted that several distinct fractional TI states can
correspond to the same $\sigma_{H,s}$. For instance,
$\sigma_{H,s}=\frac{3}{5}\frac{e^2}{2h}$ can correspond to a
$\N_f=1,$ $U(1)_\mathrm{em}\times SU(5)$ theory with
$\theta=3\pi$, or to a $\N_f=2$, $U(1)_\mathrm{em}\times
U(3)\times U(4)/U(1)$ theory with $\theta_1=\pi$ and
$\theta_2=-3\pi$. This is analogous to the well-known fact that
topological orders in FQH states cannot be characterized by the
Hall conductance alone~\cite{topdegeneracy}. The property of GSD
provides a finer classification
of fractional TI. 
Moreover, all the states discussed so far can be considered as
$(3+1)$-dimensional generalization of Abelian FQH states. In
principle, more generic non-Abelian fractional TI states with
exotic even-denominator `halved' non-Abelian FQH states on their
surface can also be constructed using more generic parton
decompositions, which correspond to effective theories with gauge
groups other than $SU(m)$~\cite{Barkeshli2009}.

In conclusion, we have shown that fractional TI states in $3+1$
dimensions with a quantized fractional bulk magnetoelectric
polarization $P_3$ and a `halved' odd-denominator surface FQHE are
fully consistent with $T$-invariance, and can in principle be
realized in strongly correlated systems with strong spin-orbit
coupling.

We thank M. Barkeshli, M. P. A. Fisher, M. Freedman, S. Kivelson,
Z. Wang, X.-G. Wen, S. Yaida, and H. Yao for helpful discussions.
J.M. is supported by the Stanford Graduate Program, X.L.Q is
supported by Microsoft Research Station Q, A.K. is supported in
part by DOE grant DE-FG02-96ER40956, and S.C.Z. is supported by
the NSF under grant numbers DMR-0904264.

After this paper was originally posted, fractional TI were further investigated by B. Swingle {\it et al.}~\cite{swingle2010}. We acknowledge the insightful discussion with the authors M. Barkeshli, J. McGreevy, and T. Senthil of Ref.~\onlinecite{swingle2010} which was very valuable for modifications to our original draft.

\bibliography{frac3dti}

\begin{thebibliography}{38}
\bibitem[{dec()}]{deconfined}
\bibinfo{note}{This can be achieved either by adding additional colored but
  electrically neutral matter or considering the special case with only Abelian
  groups, $\mathcal{N}_c^{(f)}=1$ for all $f$.}
\end{thebibliography}

\end{document}